\documentclass[aps,prb,showpacs,twocolumn,10pt]{revtex4-1}
\usepackage[english]{babel}
\usepackage[ansinew]{inputenc}
\usepackage{graphicx}
\usepackage{graphics}
\usepackage{amsmath}
\usepackage{amsfonts}
\usepackage{amssymb}
\usepackage{epstopdf}
\usepackage{makeidx}
\usepackage{subfigure}
\usepackage[usenames,dvipsnames]{color}
\usepackage{pgf}
\usepackage{bm}
\usepackage{tikz} 
\usetikzlibrary{decorations.pathmorphing,positioning,spy}
\usepackage[pdfstartpage=1]{hyperref}
\hypersetup{colorlinks=true,citecolor=blue,linkcolor=blue,filecolor=blue,urlcolor=blue}
\begin{document}
\title{Disorder assisted transmission due to charge puddles in monolayer graphene:
Transmission enhancement and local currents}
\author{Leandro R. F. Lima}
\affiliation{Instituto de F\'{i}sica,
  Universidade Federal Fluminense, 24210-346 Niter\'{o}i, Brazil}
\author{Caio H. Lewenkopf}
\affiliation{Instituto de F\'{i}sica,
  Universidade Federal Fluminense, 24210-346 Niter\'{o}i, Brazil}

\date{\today}

\begin{abstract}
We investigate the contribution of charge puddles to the non-vanishing conductivity 
minimum in disordered graphene flakes at the charge neutrality point.
For that purpose, we study systems with a geometry that suppresses the transmission 
due to evanescent modes allowing to single out the effect of charge fluctuations in the 
transport properties. 
We use the recursive Green's functions technique to obtain local and total transmissions 
through systems that mimic vanishing density of states at the charge neutrality point in
the presence of a local disordered local potential to model the charge puddles.
Our microscopic model includes electron-electron interactions via a spin resolved Hubbard 
mean field term.
We establish the relation between the charge puddle disorder potential and the electronic 
transmission at the charge neutrality point.  
We discuss the implications of our findings to high mobility graphene samples deposited 
on different substrates and provide a qualitative interpretation of recent experimental 
results. 
\end{abstract}


\maketitle
\section{Introduction}
\label{intro}

The peculiar electronic transport properties of graphene have triggered numerous 
experimental and theoretical studies \cite{castroneto09,Mucciolo10,DasSarma11}. 
Of particular interest is the conductivity of graphene single-layers at the charge 
neutrality point.
Experiments \cite{Miao07,Danneau08} have confirmed the theoretical 
prediction \cite{Tworzydlo06} that the conductivity minimum is $4e^2/\pi h$ for short 
and wide undoped ballistic samples.
For larger graphene high mobility flakes deposited on oxide substrates 
\cite{Geim07,Tan07}, the conductivity shows a minimum close to $4e^2/h$. 
Theoretical works have shown that the conductivity of graphene
at the Dirac point increases with disorder \cite{Rycerz07,Titov07,Lewenkopf08}.
This counterintuitive result is interpreted as a manifestation of Klein tunneling 
\cite{Katsnelson06,Cheianov06} and weak anti-localization \cite{Hikami80}.
Inhomogeneous electron-hole charge puddles \cite{Martin08,Zhang2009}
are believed to be the main source of disorder in undoped graphene systems 
\cite{Cheianov07,castroneto09,Katsnelson12book}.

 Charge puddles are ubiquitous in single-layer graphene samples deposited on a 
substrate  \cite{Martin08,Zhang2009,Xue11}, but their origin is still under 
debate \cite{Ponomarenko09,Couto11,Deshpande09,Gibertini12,Martin15}. 
Charge inhomogeneities can be formed, for instance, by charges trapped in 
the substrate \cite{Martin08,Rossi08}. 
Some authors argue that ripples can induce charge puddles \cite{Gibertini12,Martin15}, 
but typical experimental data show only weak spacial correlation between the 
later \cite{Deshpande09,Gibertini12}.
Transport measurement through graphene on substrates with very different dielectric 
constants \cite{Ponomarenko09,Couto11} show surprisingly similar sample mobilities, 
indicating that charge distribution fluctuations are probably not the dominant mechanism 
for electron momentum relaxation processes at high doping. 

Scanning tunneling microscopy (STM) studies \cite{Dean10,Xue11} of the 
local chemical potential $\mu_{\rm loc}$ at charge neutrality reveal that the 
charge fluctuations in graphene monolayers on hexagonal boron nitride 
(hBN)  \cite{Dean10} are about an order of magnitude smaller than those 
on silicon dioxide (SiO$_2$) samples \cite{Xue11}. In both cases the data 
show that $\mu_{\rm loc}$ follows a Gaussian distribution with a standard 
deviation of $5.4 \pm 0.1$ meV for hBN and $55.6 \pm 0.7$ meV for SiO$_2$. 
Some authors find a typical charge puddle size $a\approx 20$ nm 
\cite{Deshpande11} independent of substrate, while others \cite{Xue11} show
 that the puddles in graphene on SiO$_2$ are smaller than those of graphene 
 on hBN.

Several theoretical studies investigate the effect of a local long range local or chemical 
potential disorder at the charge neutrality point \cite{Rycerz07,Lewenkopf08,Adam09,Mucciolo10}. 
In terms of the ratio between the electron elastic mean free path $\ell$ and the system
size $L$, the following picture emerges: While for $\ell/L>1$ the transport is ballistic and dominated by evanescent modes \cite{Tworzydlo06}, deep in the diffusive regime, $\ell/L \ll 1$ the conductivity 
is governed and enhanced by potential fluctuations scattering that lead to weak 
anti-localization \cite{Mucciolo10}.
 
Recent experimental studies report an insulator behavior at the neutrality point in 
single-layer graphene on boron nitrite \cite{Amet13,Woods14}.
Ref.~\onlinecite{Woods14} shows that the conductivity minimum depends strongly on the 
matching between the graphene and the hBN lattices constants.
Ref.~\onlinecite{Amet13} observes that a resistivity as high as several megohms per 
square for low temperatures, $T \approx 20$~mK, with a power law increase with 
temperature. 
A metal-insulator transition driven by decreasing rather than by increasing the charge puddle disorder has been also reported in graphene double-layers \cite{Ponomarenko11}.
These observations do not fit in the general picture and call for further
investigation.

In summary, although it is widely accepted that disordered charge puddles are responsible 
for an enhancement of the conductivity minimum at the CNP, there is very little quantitative support for this picture, particularly at the ballistic-diffusive crossover regime.
In one hand, analytical works rely on semiclassical arguments, that require charge puddles 
with a large number of electrons \cite{Cheianov07,DasSarma11}, a condition hardly met by experiments. 
On the other hand, numerical simulations typically contain contributions of evanescent 
modes \cite{Tworzydlo06} to the transmission that are inextricably mixed with those due 
to charge inhomogeneities, obscuring the later. 
The main goal of this paper is to disentangle these contributions and to single out the effects 
of charge puddles in the conductivity of disordered graphene sheets close to the neutrality 
point.

For that purpose, we revisit this problem and analyze the transport properties using a 
self-consistent recursive Green's functions (RGF) technique \cite{Areshkin09,Lewenkopf13} 
with spin resolution that includes the electronic interaction via a mean field Hubbard term.
We calculate electronic current densities between neighboring carbon sites.
We analyze the electronic propagation near {\it pn} charge puddle interfaces,
relating the general transport properties to the typical puddles characteristics, such as 
their charge, size and shape.

This paper is organized as follows.
In Sec.~\ref{sec:model} we present the model Hamiltonian we employ to describe 
graphene sheets with disordered charge puddles.
There we also discuss the key ingredients necessary to calculate transport properties 
and to realistically assess the minimum conductivity at the CNP using a lattice model 
of a moderate size.
In Sec.~\ref{sec:results} we present the total and local transmissions for different 
potential profiles, establishing a qualitative understanding of the role of charge puddles 
in the electronic transport.
We present our conclusions in Sec.~\ref{sec:conclusions}.

\section{Model and theory}
\label{sec:model}

We model the electronic properties of a monolayer graphene sheet by a Hubbard mean field 
$\pi$-orbital tight-binding model, namely \cite{castroneto09,Rossier07,Areshkin09}
\begin{align}
	H = & -t\sum_{\left\langle i, j\right\rangle, \sigma} a^\dagger_{i,\sigma} a_{j,\sigma}
		+  \sum_{i,\sigma} U\left(\left\langle \hat n_{i,-\sigma}\right\rangle-\frac{1}{2}\right) 	\hat n_{i,\sigma} &
		\nonumber\\
	& +  \sum_{i,\sigma} V_i \,\hat n_{i,\sigma},
\label{eq:HMF}
\end{align}
where $a^\dagger_{i,\sigma}$ ($a_{i,\sigma}$) stands for the operator that 
creates (annihilates) an electron with spin $\sigma$ at the site $i$, 
$\hat n_{i,\sigma}$  is the corresponding electron number operator, while 
$\left\langle \hat n_{i,\sigma}\right\rangle$ is its expectation value, $t$ is the 
hopping matrix element connecting states at neighboring sites, and 
$\left\langle i, j\right\rangle$ indicates that the sums are restricted to first 
neighbors sites.  The electron-electron interaction is approximated by the 
Hubbard mean field term, where $U$ is the Coulomb energy for double 
occupancy of a carbon site \cite{Areshkin09}.

We assume that the electron-hole puddles are generated by a disordered 
long-range local potential  $V(\mathbf r)$.
We model $V(\mathbf r)$ in the lattice, $V_i=V(\mathbf r_i)$, by a superposition of 
$N_G$ Gaussian potentials centered at the positions $\mathbf r_p = (x_p,y_p)$, 
namely
\begin{align}
V_i = \sum_{p=1}^{N_G} V_p\ \text{exp}\left[-2\frac{(x_i-x_p)^2}{d_x^2}-2\frac{(y_i-y_p)^2}{d_y^2}\right].
\label{potential}
\end{align}
We consider $V_p$ to be either $V$ or $-V$ with equal probability and the 
positions ${\mathbf r}_i$ are random and uniformly distributed over the 
graphene flake.
Suitable choices of the Gaussian range parameters $d_x$ and $d_y$ allow 
us to study different physical regimes, as discussed in the next Section. 
For $d_x=d_y$ this model is equivalent to a Gaussian disordered model studied 
by several authors \cite{Rycerz07,Lewenkopf08,Adam09}.

\subsection{Model geometry}\label{modelgeometry}

\begin{figure}[htbp]
	\centering
		\includegraphics[width=1.00\columnwidth]{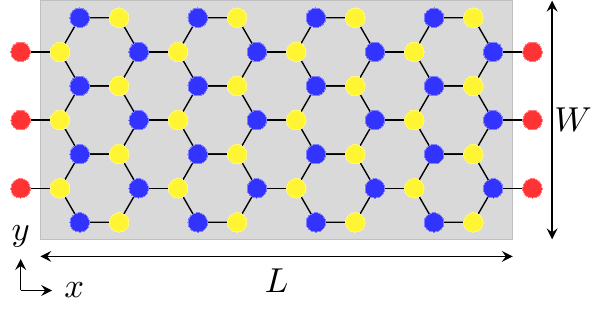}
	\caption{(color online) Sketch of an armchair GNR of length $L$ and width $W$. 
The red sites represent semi-infinite linear chains connected to source or drain reservoirs.
}
\label{fig:ribbon}
\end{figure}

Our main goal is the study of the effect of charge puddles in transport properties in 
graphene flakes near the charge neutrality point. From the perspective of simulations,
the difficulty is that the current numerical methods based on microscopic 
models that take into account interactions or address local transport properties 
are computationally prohibitive for systems of realistic sizes. For that reason we 
study much smaller systems, with similar properties of bulk graphene and 
resort to a scaling scheme to draw conclusions.  In what follows we show that this 
is accomplished by using armchair graphene nanoribbons (GNRs), as the one 
depicted in Fig.~\ref{fig:ribbon}. 

In the absence of electron-electron interaction, GNRs with zigzag edges are always 
metallic. Armchair GNRs are metallic when the number of hexagons across the transverse 
direction is $M=3i$, where $i$ is an integer number, and semi-conductor otherwise 
\cite{Brey2006}.
Both zigzag and armchair metallic GNRs show an unit transmission per spin channel 
as the doping goes to zero.
This metallic behavior, related to boundary effects, is unlikely to be manifest in the bulk 
and makes difficult to single out the effects of charge puddles in the conductance.
Electronic interactions give rise to a gap in zigzag and chiral GNRs with pristine edges, 
which is an condition hardly met by graphene flakes.
Thus, we find more convenient to use semi-conductor armchair GNRs for this study.

It is convenient to express the length $L$ and the width $W$ of armchair GNRs as 
$L=N(\sqrt{3}a_0/4)$ and $W=Ma_0$, where $N$ gives the number of sites in an 
armchair chain along the GNR length and $M$ is the number of hexagons across 
its width. $a_0=2.46$ \AA \ is the carbon-carbon bond length. The total number of 
sites in the system is $N_{\rm tot} = (2M+1)N/2$. Figure \ref{fig:ribbon} illustrates 
an armchair GNR with $M=3, N=16$ and $N_{\rm tot} = 56$.
 
The energy threshold $E_1$ to open the first propagating channel depends 
on the nanoribbon width $W$ (or equivalently on $M$) roughly as \footnote{We estimate that $E_1 \approx a_0t/W$ from the analysis of $E_1$ versus 
$W^{-1}$ obtained from the numerical solution of the tight-binding model  for several semiconductor pristine armchair-edge graphene strips of widths up to 
$W=160a_0$ (a similar analysis as that in Ref.\onlinecite{Brey2006}).
} $W^{-1}$.
With increasing $W$, the system properties become increasingly similar to bulk 
graphene: the bands collapse into a conical one and the energy gap goes to zero, 
resulting in a  vanishing DOS at the charge neutrality point.

\subsection{Transport calculations}

{
We obtain the system transport properties using the non-equilibrium Green's 
function technique \cite{Meir92,Haug08}. 
We calculate the system Green's functions using the mean field Hamiltonian given by 
Eq.~\eqref{eq:HMF} for the two-contact lattice geometries shown in Fig.~\ref{fig:ribbon}.
For that purpose, we use the recursive Green's function method \cite{Lewenkopf13} 
combined with a self-consistent procedure \cite{Areshkin09} that we describe below.
}

We compute the conductance using the partition geometry shown in Fig.~\ref{fig:ribbon}. 
The semi-infinite chains placed at the right (R) and the left (L) side of the central 
region represent the leads that connect the graphene flake to source and drain 
reservoirs. 
The system Hamiltonian, Eq.~\eqref{eq:HMF}, the Green's 
functions, and the transport properties depend self-consistently on the system electronic
density $\langle \hat{\mathbf n}\rangle$.

In the zero bias limit, the self-consistent relation that connects the electronic density 
and the system retarded Green's function $\mathbf G^r$ reads \cite{Haug08}
\begin{align}
\langle \hat n_{i,\sigma}(\mu) \rangle = -\int_{-\infty}^{\infty} 
\frac{dE'}{\pi} f(E')\ \text{Im}\,\left[ G^{r,\sigma\sigma}_{i,i}(E') \right],
\label{equilibrium-density}
\end{align}
where $f(E)$ is the Fermi-Dirac distribution at the source and drain reservoirs with 
corresponding chemical potentials $\mu_{\rm L} \approx \mu_{\rm R} \approx \mu$.

For the systems of interest, where $N_{\rm tot}\gg 1$, $\mathbf G^r(E)$ has a large number of 
complex poles and shows fast energy variations close to charge neutrality. 
Thus, a real-axis numerical evaluation of the integral in Eq.~\eqref{equilibrium-density} is 
very costly, since a good accuracy demands a fine integration mesh. 

Efficient methods \cite{Ozaki07,Croy09,Areshkin10} developed to evaluate the integral in 
Eq.~\eqref{equilibrium-density} use complex analysis, taking advantage of the analytical 
structure of $\mathbf G^r(E)$. 
Since the poles $\varepsilon_p$ of retarded Green's function lie in the lower complex half-plane, 
Im$(\varepsilon_p)<0$, Eq.~\eqref{equilibrium-density} is readily evaluated by a contour integration.
The integration limits have to be treated carefully: To guarantee charge conservation all states 
must be inside the integration limits \cite{Areshkin10}. 
The Fermi function provides an effective upper energy cut-off, but introduces Matsubara poles 
in the upper complex half-plane. 
To efficiently deal with these issues, we use the integration technique described by Ozaki in 
Ref.~\onlinecite{Ozaki07}. 
The method expands the Fermi-Dirac distribution in a partial 
fraction decomposition, so that the integral in Eq.~\eqref{equilibrium-density} 
is given by a sum of $N_{\bar E}$ evaluations of $\mathbf G^{r}(\bar E_j)$ at the complex energies 
$\bar E_j$, with $j=1,\cdots,N_{\bar E}$. \cite{Ozaki07}

Two key features of this method are noteworthy: 
(i) There is no need to specify the lower energy bound and (ii) the integration 
precision is controlled by 
varying $N_{\bar E}$.
To attain a given accuracy, smaller temperatures require larger $N_{\bar E}$ values.
We choose $kT = 25$ meV. This temperature is very amenable for the numerical calculation and, since $kT/E_1<1$, it still guarantees that we address a low temperature regime for the systems we study.
We find that $N_{\bar E} \approx 40-50$  guarantees an error smaller than $10^{-5}$ for the 
electronic densities we study in this paper. 
For the systems we analyze, this method is three orders of magnitude faster than a 
real-axis integration.  
Nonetheless, the calculation of $\langle \hat{\mathbf n}\rangle$ still remains as the computational bottleneck that 
limits the conductance evaluation of large model systems. 

The self-consistent procedure we employ is rather standard: We start with an 
initial guess for $\langle \hat{\mathbf n}_{\rm in}(\mu)\rangle$, obtain the system 
retarded Green's function $\mathbf G^r(E)$ using the RGF method \cite{Lewenkopf13}, 
and calculate the updated equilibrium electronic density 
$\langle \hat{\mathbf n}_{\rm out}(\mu)\rangle$ using Eq.~(\ref{equilibrium-density}).
For the subsequent iterations we use the modified second Broyden method 
\cite{Singh86,Ihnatsenka07,Areshkin10}, that mixes all the previous input and 
output electronic densities to construct an optimized input 
$\langle \hat{\mathbf n}_{\rm{in}}(\mu)\rangle$ 
for the next self-consistent iteration. 
The procedure is repeated until convergence is achieved. 
Our convergence criteria is  $\left|\langle \hat{\mathbf n}_{\rm{in}}(\mu)\rangle-
\langle \hat{\mathbf n}_{\rm{out}}(\mu)\rangle\right|  < 10^{-5}$, 
for which the required number iterations is 20 up to $\sim$40 depending on 
system size. 
For the systems we study, self-consistent loop procedures that naively 
update $\langle \hat{\mathbf n}_{\rm{in}}(\mu)\rangle$ with the occupations 
obtained from the previous iteration, $\langle \hat{\mathbf n}_{\rm{out}}(\mu)\rangle$, 
are about $10^2$ times slower than those that use the Broyden method. 
\footnote{
When calculating the electronic properties as a function of $\mu$, the number 
of iterations can be further diminished by using  as the initial guess for 
$\mu + \Delta \mu$ the converged electronic density obtained for $\mu$. 
In this case, one has to find an optimal compromise between the number of 
self-consistent iterations and the chemical potential steps $\Delta \mu$.}

Once convergence is achieved, we calculate the transport quantities,
such as the total transmission coefficient between L and R contacts \cite{Meir92}
\begin{align}
T_{\rm {L,R}}^{\sigma}(E) = \text{Tr}\left[
\mathbf \Gamma_{\rm L}^{\sigma}(E) \mathbf G^{r,\sigma\sigma}(E) 
\mathbf \Gamma_{\rm R}^{\sigma}(E) \mathbf G^{a,\sigma\sigma}(E)\right]
\label{total-transmission}
\end{align}
and the local transmission coefficient between the neighboring $i$ and $j$ sites 
\cite{Nonoyama98,Cresti03,Nikolic06,Zarbo07,Lewenkopf13} 
\begin{align}
T_{i,j}^{\sigma}(E) = -2t\,\text{Im} \left\{ \left[ 
        \mathbf G^{r,\sigma\sigma}(E) \mathbf \Gamma_{\rm L}^{\sigma}(E) \mathbf G^{a,\sigma\sigma}(E) \right]_{i,j}  \right\}.
\label{local-transmission}
\end{align}
The advanced Green's function is obtained from $\mathbf G^{a,\sigma'\sigma}(E) = 
[\mathbf G^{r,\sigma\sigma'}(E)]^\dagger$.
The line width functions are  $\mathbf \Gamma_\alpha^{\sigma}(E) = 
-2 \,\mbox{Im}\,\mathbf \Sigma^{r,\sigma}_\alpha(E)$.
Here $\mathbf\Sigma^{r,\sigma}_\alpha$ is the retarded self-energy associated 
to the decay into $\alpha =$ L and R leads and is calculated following a standard 
procedure \cite{Lewenkopf13}. 

Equations (\ref{total-transmission}) and (\ref{local-transmission}) assume that 
the injection of electrons is spin independent, $\mathbf \Gamma_\alpha^{\sigma} = 
\mathbf \Gamma_\alpha^{\bar\sigma}$, and the absence of spin-flip processes.
Thus, at sufficiently low temperatures, the zero bias limit conductance of the system 
for an electronic energy $E$ is $G(E) = 2 (e^2/h) T_{\rm{L,R}}(E)$, where 
$T_{\rm{L,R}} \equiv T_{\rm{L,R}}^\sigma = T_{\rm{L,R}}^{\bar\sigma}$.
In the diffusive regime, the conductance $G$ can be converted into a conductivity 
$\sigma$ using $\sigma = (L/W) G$.

In the absence of electron-electron interactions, the aspect ratio $L/W$ dictates 
the conductance of pristine graphene sheets \cite{Tworzydlo06}. For $W/L \gg 1$, 
evanescent modes lead to a conductivity minimum of the order $e^2/h$ at the 
charge neutrality point, both for semiconductor and metallic graphene ribbons 
\cite{Tworzydlo06}. 
In the opposite limit of narrow and/or long ribbons, $W/L \ll 1$, the conductivity 
goes to zero.

\begin{figure}[!htbp]
\centering
\includegraphics[width=0.95\columnwidth]{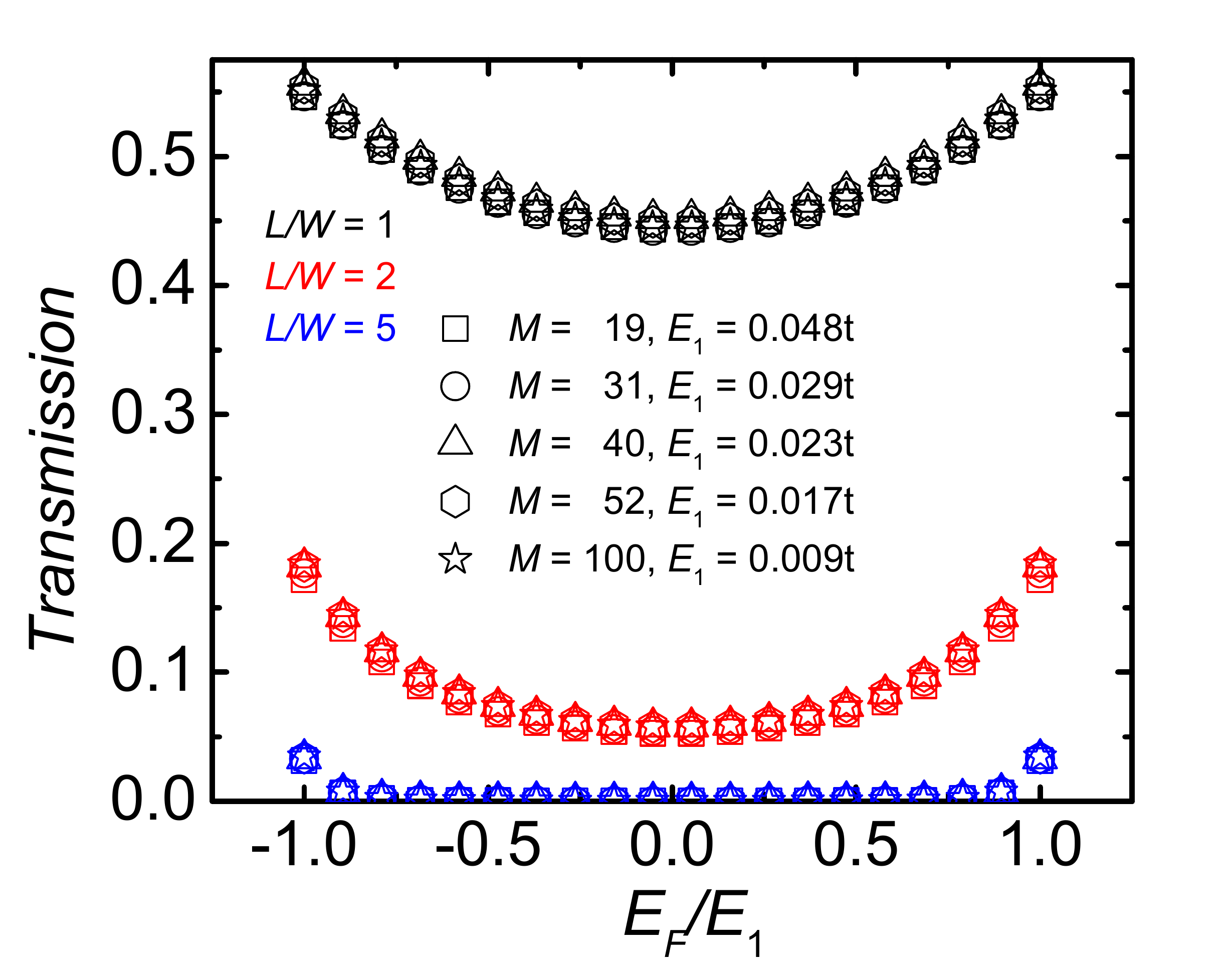}
\caption{(color online) Total transmission $T$ of armchair GNRs of different widths $W$ 
connected to generic metallic leads as a function of the ratio between the energy $E_F$ 
and first threshold energy of first conducting channel $E_1(W)$. The behavior of $T$ versus
$E_F/E_1$ depends only on the GNR aspect ratio $L/W$. 
}
\label{fig:ratio_collapse}
\end{figure}

We find that, for ballistic graphene systems, electron-electron interactions do not 
qualitatively change this picture.
This statement is based on the study of the transmission through pristine semiconductor 
graphene flakes with armchair edges connected to generic metallic leads. We consider 
different sizes and aspect ratios using $U=t$ \cite{Schuler2013}.
Our results are summarized in Fig.~\ref{fig:ratio_collapse}.

As is the non-interacting case, the transmission dependence on the geometry can be 
cast in terms of the aspect ratio. We compute $T(E_F)$ for $L/W=1,2$ and $5$. 
Since we work in the low temperature regime, we define $E_F=\mu$.
For each value of $L/W$ we consider different system sizes defined by $M$. 
Recall that $W=Ma_0$.
Figure \ref{fig:ratio_collapse} shows that by expressing the electronic energy as 
$\varepsilon = E_F/E_1$, all $T(E_F)$ corresponding to GNRs with the same aspect 
ratio $L/W$ collapse into a single curve.
For $L/W=1$, the transmission minimum at $E_F=0$ is roughly $T\approx 0.44$.
Electron-electron interaction effects in the mean field approximation only affect 
strongly the transmission for states with high density of states at the edges of 
zigzag GNRs \cite{Carvalho2014,Son2006PRL}, which is not the case of armchair 
GNRs, the non-interacting value of the transmission minimum is roughly the same 
as that for $U=0$.

The non-interacting result $T(0)=2/\pi \approx 0.64 $ found analytically in 
Ref.~\onlinecite{Tworzydlo06} and reproduced numerically in Ref.~\onlinecite{Lewenkopf13} 
differs from our calculation due to the different modeling of the leads. 
While we use linear chain as contacts, these previous works used square lattices that provide additional non-diagonal self energy elements, changing the leaking probability of the electrons.
For $L/W=2$, the transmission minimum decreases to about $0.05$.
Finally, zero transmission is obtained if the ratio is as large as $L/W=5$.

These results show that the evanescent modes contribution indeed depends only 
on the aspect ratio $L/W$ and that their contribution to the conductance is almost 
entirely suppressed for $L/W>5$.
As a consequence, even a $5\ \mu$m long semiconductor graphene flake may have a 
non-vanishing transmission minimum at $E_F=0$ due to evanescent modes if 
$W>1\ \mu$m. 
In the remaining of this paper we eliminate the effect of evanescent modes by considering 
graphene systems with $L/W\ge 5$.

\section{Results}
\label{sec:results}

In this section we study the effects of charge puddles on the transmission minimum 
close to the charge neutrality point by considering different models for $V(\mathbf r)$. 
To develop some insight on the role of interactions and the variations on the local 
potential, we begin discussing the simple case of a $pn$-junction before
we proceed to cases of disordered charge puddles.

\subsection{\texorpdfstring{\textit{pn}}{pn} junctions: 
                    \texorpdfstring{$U=0$}{U.EQ.0} limit case}
\label{pn}

We model the {\it pn}-junction interface potential $V(\mathbf r)$ by taking $d_y \gg W$ in 
Eq.~(\ref{potential}), that corresponds to a constant potential along the GNR width.
{
We consider a system with $M=52$ and choose the smallest value of $N$ that gives an 
aspect ratio  $L/W \ge 5$, namely $N=604$. 
We generate a \textit{pn} junction by placing a positive Gaussian potential centered 
at the site $(L/4,W/2)$ and a negative one at $(3L/4,W/2)$.
We choose $d_x = 0.24L$.
}
At the $p$ and $n$-doped regions, the local potential is constant and set to 
$V({\bf r}) = V$ for $0\le x\le L/4$ and $V({\bf r}) = -V$ for $3L/4\le x\le L$.
This choice renders a $V({\bf r})$ with a smooth Gaussian transition between 
positive and negative doped regions.

\begin{figure}[!htbp]
\centering
\includegraphics[width=\columnwidth]{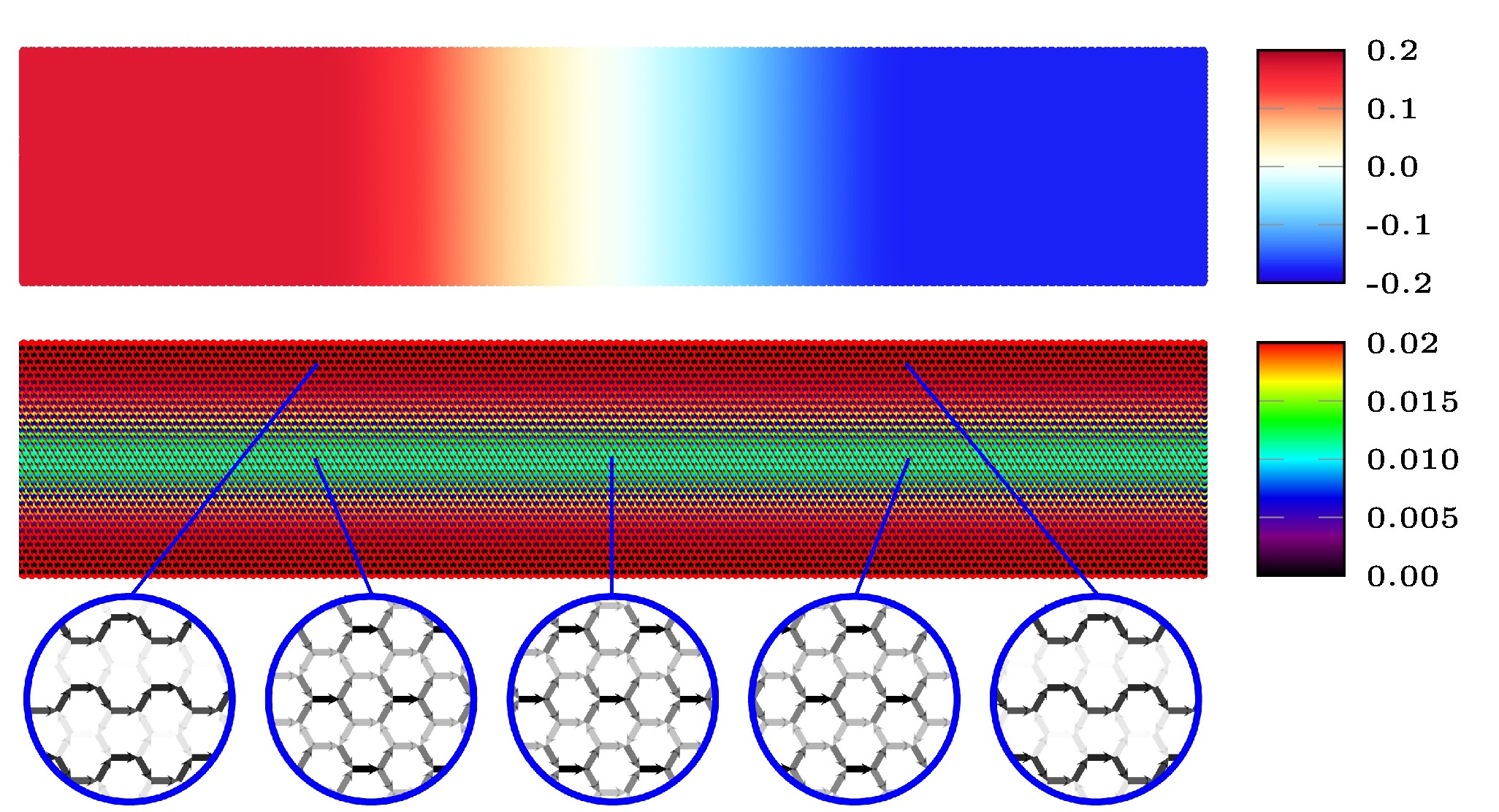} 
\caption{
(color online) Top: Potential profile of a smooth \textit{pn} junction for $M = 52$. 
The energy scale on the right is given in units of $t$. 
Middle: Corresponding profile of the local transmission at $E_F\approx 0.15E_1$.
Bottom: Zoom of the local transmission at selected areas.
}
\label{fig:999}
\end{figure}

\begin{figure}[!tbp]
\centering
\includegraphics[width=0.95\columnwidth]{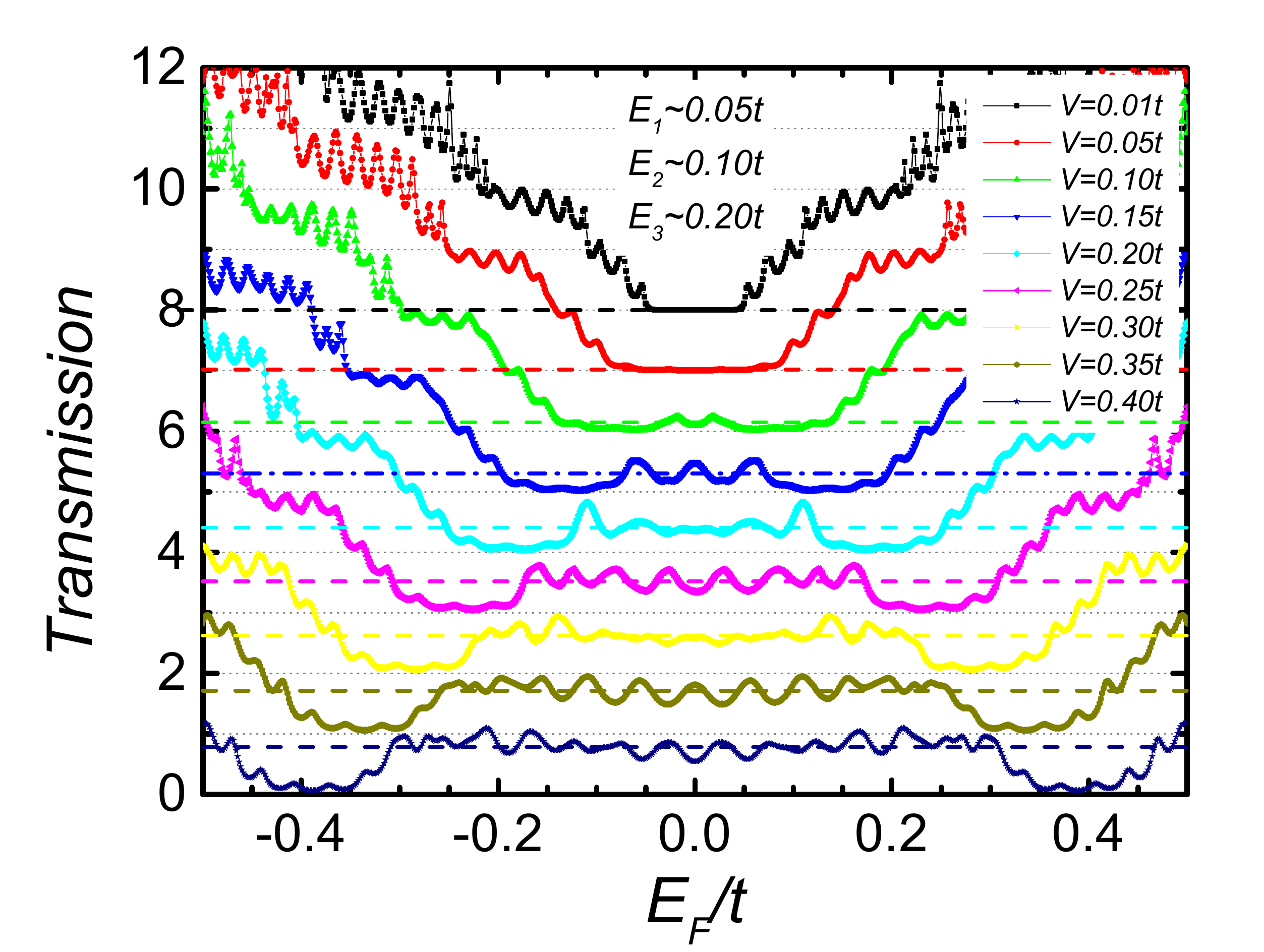}
\caption{(color online) Transmission as a function of the electronic energy for the 
\textit{pn} junction shown in Fig.~\ref{fig:999} for different potential values from 
$V=0.01t$ through $V=0.40t$. Here $M=19$ and $U=0$. 
The dashed lines correspond to $w_{\rm tun}$, Eq.~(\ref{tunneling_falko}), 
for each value of $V$. 
For clarity, the transmissions are shifted by $T=1$ for successive values of $V$.
}
\label{fig:transmission_pn_offset}
\end{figure}

{
We obtain $E_1$ by inspecting the corresponding dispersion relation. 
We choose the potential strength $V=10E_1$. The top panel of Fig.~\ref{fig:999} 
shows the contour plot of the ${\it pn}$-junction potential $V({\bf r})$, while 
Fig.~\ref{fig:transmission_pn_offset} shows $V({\bf r})$ along the system 
longitudinal direction.
}

We compute the local transmissions for selected energies close to the charge 
neutrality point using Eq.~(\ref{local-transmission}). 
We recall that in this subsection we set $U=0$.
The middle panel of Fig.~\ref{fig:999} shows the local current density profile 
at the energy $E_F=0.15E_1$.
This energy lies inside the transport gap of the GNR in the absence of the 
{\it pn}-junction, that is. for  $V=0$.
The current near the system edges is mainly transmitted through ``armchair chains'' 
(see insets at the bottom of Fig.~\ref{fig:999}), similar to the transmission through 
pristine armchair GNRs.
At the center of the ribbon, backscattering processes mix the transmission between 
different armchair chains generating a richer transmission structure.
The local transmission profile is almost invariant along the system longitudinal direction, 
including the $p$ and $n$ doped regions and their interface.

Figure~\ref{fig:transmission_pn_offset} shows the total transmission $T(E_F)$ as a 
function of the electronic energy $E_F$ for several values of $V$.
The main features are: 
(i) $T(E_F)$ shows Fabry-Perot interference oscillations caused by backscattering at the abrupt 
potential interfaces between graphene central region and the right and left contacts.
(ii) The transport gap, centered at $E_F=0$ for $V=0$, appears twice at $E_F+V$ 
and at $E_F-V$. 
(iii) Around the CNP, between the two gaped regions, where otherwise one would 
expect a transport gap, the transmission increases with $V$.
(iv) For large values of $V$ (see, for instance, $V/E_1 \gtrsim 7$), the transmission 
near the CNP shows $E_F$ dependent fluctuations, but $T(E_F)$ remains between $0.5$ and $1$.

\begin{figure}[!htbp]
\centering
\includegraphics[width=\columnwidth]{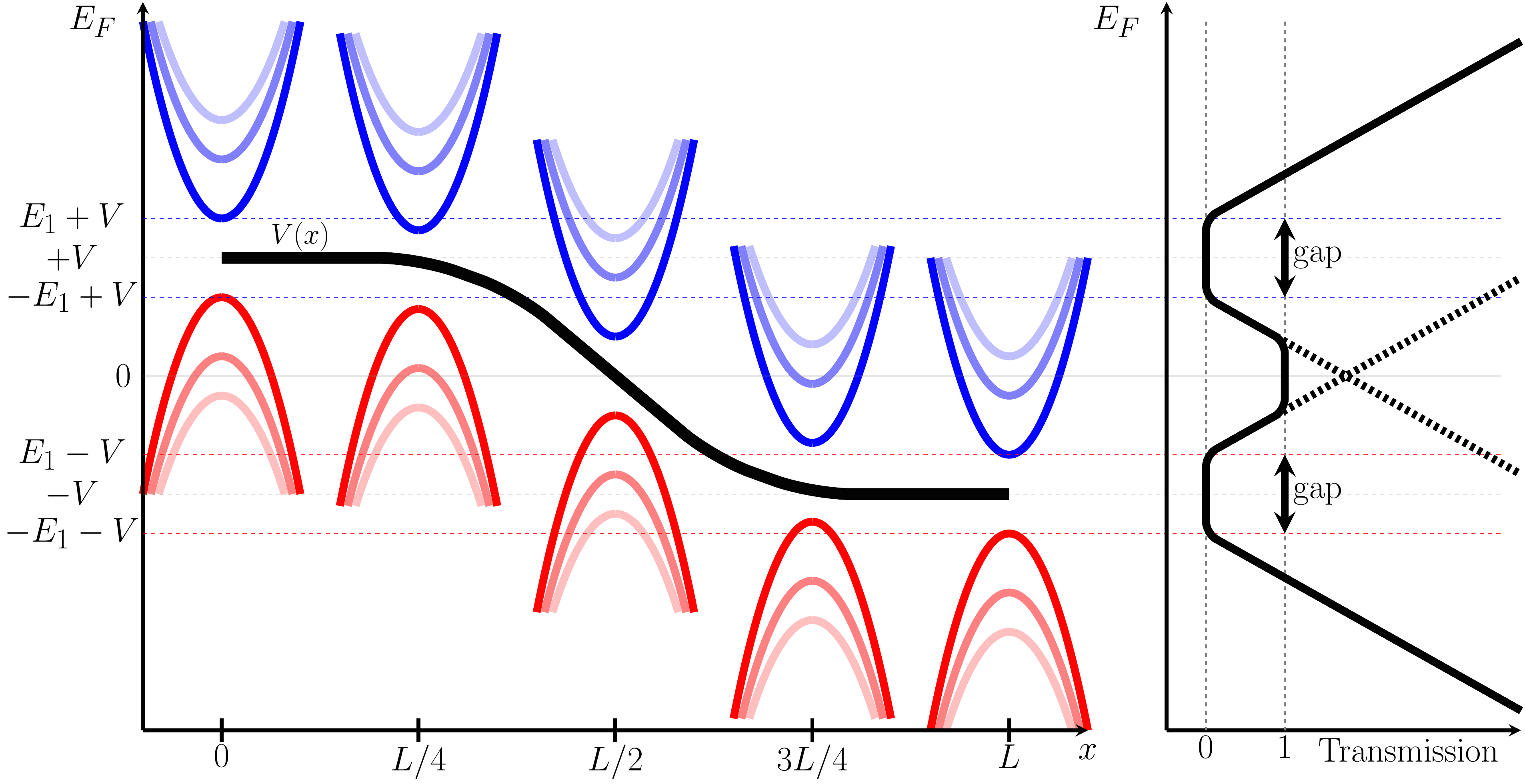}
\caption{(color online) Left: Sketch of the local dispersion relation at five representative 
points $x=0, L/4, L/2, 3L/4$, and $L$. The black line along the length $L$ represent 
the potential profile $V({\bf r})$ due to a smooth \textit{pn} junction.
The dashed gray lines at $+V$ and $-V$ indicate the local CNP for $x=0$ and $x=L$, respectively. The blue (red) solid lines at $V+E_1$ and $V-E_1$ ($-V+E_1$ and $-V-E_1$) 
stand for the energies to open the first channel in the presence of the potential $V$ at $x=0$ ($x=L$).
Right: Sketch of the transmission (solid line) as a function of $E_F$ for the potential in the left 
panel. The dashed lines correspond to the undoped system transmissions, shifted 
up or down due to $V$. }
\label{fig:sketch_smooth_pn_new}
\end{figure}

In Fig.~\ref{fig:sketch_smooth_pn_new} we present a sketch that suggests a simple
explanation of the main features of the transmission through the smooth \textit{pn} 
junction in terms of the local band structure.
For $0<x<L/4$, the potential $V({\bf r})=V$ shifts the local CNP to $+V$. Similarly, 
for $3L/4<x<L$, the CNP is shifted down to $-V$. The transition from $+V$ to $-V$ 
happens in the region where $L/4<x<3L/4$.
Thus, $V({\bf r})$ leads to band gaps {$V-E_1<E_F<V+E_1$ and $-V-E_1<E_F<-V+E_1$} at
the~``left" and~``right" sides of the junction, respectively. 
This is illustrated by the right panel of Fig.~\ref{fig:sketch_smooth_pn_new}.
For energies around the global CNP ($E_F=0$), that is, for $-V+E_1<E_F<V-E_1$, the 
available states at $x=0$ the electrons must tunnel through at least one locally gaped 
region (at $x=L/2$).

The following picture emerges: For {$V<E_1$} the transmission is suppressed for 
{$|E_F|<V+E_1$} due the band gaps either at the right or at the left. (In our 
calculations $T$ is small but non-zero because we work with a finite $L$.) This 
corresponds to the cases where $V=0.01t$ and $V=0.05t$ shown in 
Fig.~\ref{fig:transmission_pn_offset}. { For $V>E_1$ and $|E_F|<V-E_1$, Klein 
tunneling at the {\it pn} interface dominates the transmission and the gaps appear 
only for increasing $|E_F|$.}
This transmission profile corresponds to the cases where $V>E_1$ in Fig.~\ref{fig:transmission_pn_offset} and is qualitatively captured by the sketch presented 
in the right panel of Fig.~\ref{fig:sketch_smooth_pn_new}.

Let us now estimate the magnitude of transmission at the CNP. 
For that purpose we adapt the semiclassical analysis of the Klein tunneling transmission 
presented in Ref.~\onlinecite{Cheianov06} to our case.
First, we relate local longitudinal wave number of the $n$th band with the electron energy 
$E_F$ in the presence of a \textit{pn} junction potential profile $u(x)$, as 
\begin{equation}
E_F=v\sqrt{k_x^2(E_F,x)+\left(E_n/v\right)^2}+u(x),
\end{equation}
where $v=ta_0$ and $E_n$ is the threshold energy to open the $n$th channel 
for $u(x)=0$. 
For the sake of simplicity, we approximate the \textit{pn} junction potential profile 
to $u(x) \approx Fx$, where $F=-(2V/d)x$. At the charge neutrality point, where $E_F=0$, 
the longitudinal momentum becomes $k_n(x) = v^{-1}\sqrt{F^2x^2-E_n^2}$.
In this situation, the classically forbidden region corresponds to $\ell_-<x<\ell_+$, 
where $\ell_\pm=\pm E_n/F$.
The probability of an electron in channel $n$ to tunnel through the classically 
forbidden region (gaped region) can be approximated by $w_n \approx 
\exp\left[i\int_{\ell_-}^{\ell_+}k(x)dx\right]=\exp\left(-\frac{\pi}{2ta_0}
\frac{dE_n^2}{V}\right)$.

Thus, the total tunneling probability reads 
\begin{align}
	w_{\text{tun}} \approx \sum_{n=1}^{\infty} w_n = \sum_{n=1}^{\infty} 
	\exp\left(-\frac{\pi}{4ta_0}\frac{LE_n^2}{V}\right).
\label{tunneling_falko}
\end{align}
Here we set $d=L/2$, since in our model the \textit{pn} junction potential $u(x)$ varies from 
$V$ to $-V$ in the interval $L/4<x<3L/4$.
The contribution of each channel $n$ to the tunneling probability in 
Eq.~(\ref{tunneling_falko}) decays exponentially with 
$E_n^2 \propto M^{-2}$.  
For $M$ very large, many channels contribute to the transmission and we can transform 
the sum in Eq.~\eqref{tunneling_falko} into an integral, recovering the results of 
Ref.~\onlinecite{Cheianov06}.

For the system studied in Fig.~\ref{fig:transmission_pn_offset} 
($L=220\sqrt{3}a_0/4 = 234.35$ \AA\ and $L/W=5$), we obtain $w_{\rm tun}$ to 
a good accuracy by summing over a small number of channels, $n \le N_{\text{ch}}=3$.
The ribbon band structure renders $E_1 \approx 0.05t$, $E_2 \approx 0.10t$, 
and $E_3 \approx 0.20t$.
Around the CNP the analytical transmission $w_{\rm tun}$ (dashed lines in 
Fig.~\ref{fig:transmission_pn_offset}) is in nice qualitative agreement with 
the numerical calculated one. 
We attribute the small deviations to Fabry-Perot interference patterns due the the 
wave function mismatch at the graphene-contact interface.

These observations allows us to infer the behavior of the conductance as one 
increases $W$ to realistic sample sizes: 
(i) the transmission for $|E_F|<V$ increases since the transverse mode energies 
$E_n$ decrease with $W$ and more transverse modes contribute to the transmission, 
see Eq.~(\ref{tunneling_falko}). 
(ii) the ``satellite'' gaps at $E_F=\pm V$ shrink and tend to disappear, since $E_1$ 
scales with $W^{-1}$.
(iii) The behavior for the homopolar junctions, $|E_F|>V$, remains qualitatively 
the same.  
(iv) Finally, the magnitude of the Fabry-Perot oscillations depends on the 
nature of the graphene-contact interface and will be suppressed as the ratio between 
electron-impurity mean free path $\ell_{\rm imp}$ and the system size becomes 
smaller than unit, a situation that calls for an analysis in the lines of Ref.~\onlinecite{Fogler08}.

\subsection{\texorpdfstring{\textit{pn}}{pn} junctions: \texorpdfstring{$U\neq 0$}{U.NE.0} case}
\label{gaussianpn}

We now switch on the interaction $U=t$.
The model potential is also slightly modified.
We place a positive Gaussian potential centered at $(L/4,W/2)$ and a negative 
one at $(3L/4,W/2)$, keeping $d_y \gg W$ and $d_x = 0.24L$.
This parametrization introduces a smooth variation of the \textit{pn} junction potential 
close to the contacts.

\begin{figure}[!htbp]
	\centering
		\includegraphics[width=0.95\columnwidth]{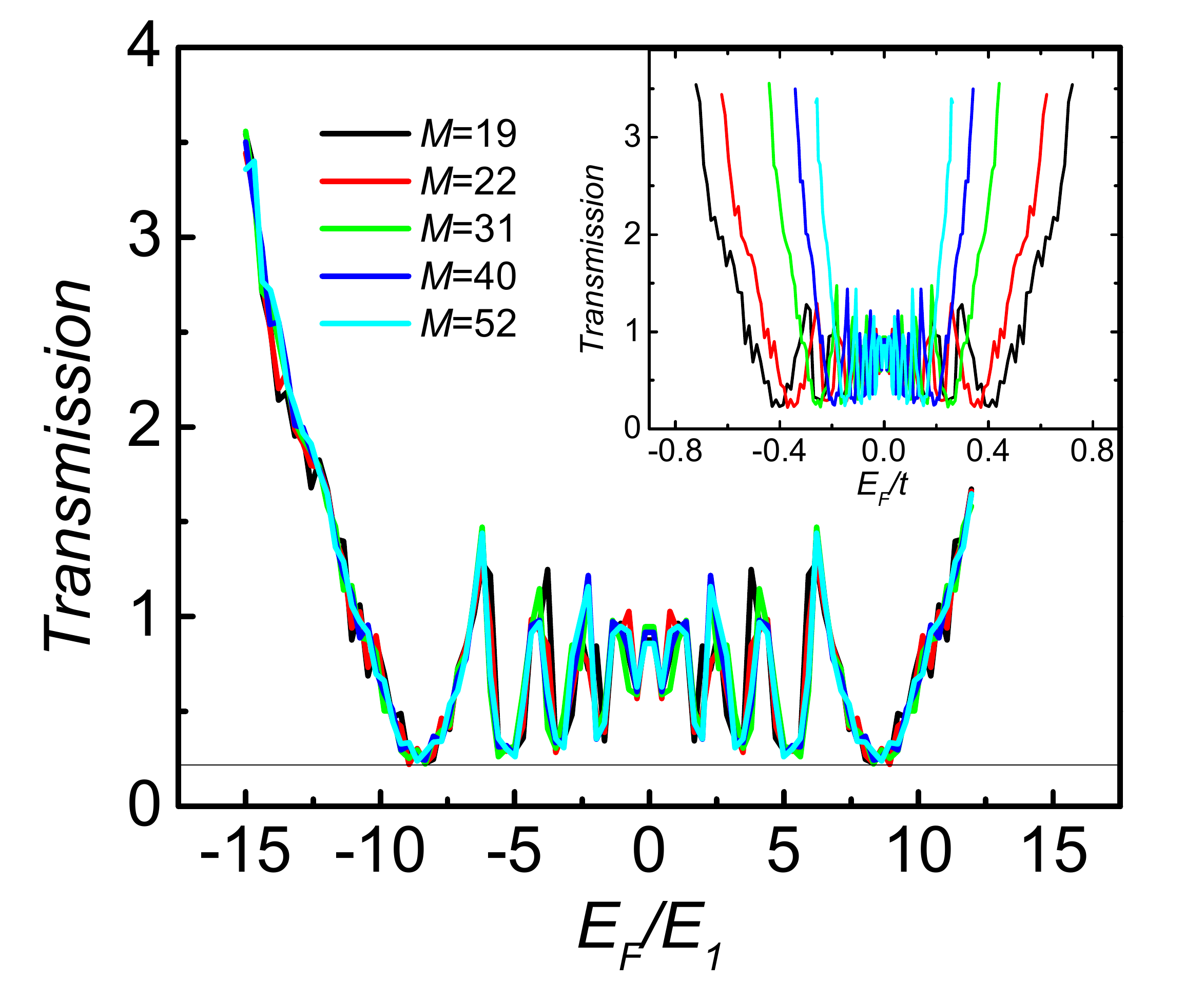}
\caption{(color online) Transmission as a function of the electronic energy $E_F$ 
in units of $E_1$ for armchair edge ribbons of different widths calculated for a \textit{pn} 
junction with a Gaussian profile. Inset: Same plot for energies in units of $t$. The solid line 
indicates the global transmission minimum $T_{\rm min}$.
	}
	\label{fig:disorder_collapse_999_inset}
\end{figure}

Figure~\ref{fig:disorder_collapse_999_inset} shows the total transmission $T(E_F)$ as a 
function of the Fermi energy $E_F$ for several values of $M$.
Here, $V=10E_1$.
Since $V({\bf r})$ is no longer constant neither for $0<x<L/4$ nor for $3L/4<x<L$,
there is no band gap alignment in these regions, which facilitates the electronic transport. 
As a result, for $V-E_1<E<V+E_1$ and $-V-E_1<E<-V+E_1$, the transmission is non-zero, in 
distinction to the case analyzed in the preceeding subsection.
For $|E_F|<V$, the transmission $T(E_F)$ displays stronger oscillations than those 
of the previous case, 
Near the CNP, the transmission remains nearly unit.

We find that by rescaling the energy $E_F$ as $\epsilon=E_F/E_1$ the transmission 
calculated for different values of $M$ collapse into a single curve. This is illustrated in 
Fig.~\ref{fig:disorder_collapse_999_inset}, inset and main panel.
For $|E_F|<V$, the electron backscattering amplitude is appreciable and its 
interference with the transmission process gives rise to the oscillating pattern 
in Fig.~\ref{fig:disorder_collapse_999_inset}.
For $|E_F|>V$ the back scattering amplitude becomes weaker and the interference 
effects are suppressed.
Like in the $U=0$, the main features of the transmission can be qualitatively explained 
by Fabry-Perot interference and Klein tunneling. 

In order to understand the onset of the  transmission minimum we also studied 
(not shown here) the transmission for several ribbon widths, and potential 
strengths $V=5E_1$, $V=10E_1$, $V=15E_1$ and $V=20E_1$. We find that:
(i) $V$ determines the energy window characterised by large interference oscillations, 
namely, $|E_F|<V$.
(ii) The transmission has an overall non-vanishing minimum, $T_{\rm min}$.
(iii) $T_{\rm min}$ does not show a simple dependence on $V$. $T_{\rm min}$ 
increases with $V$ until it saturates at a value of order of $e^2/h$. 
(iv) Most importantly, we conclude that deviations from a flat $V({\bf r})$
close to the graphene-contact interface increase the transmission, $T_{\rm min}>0$,
and eliminate the energy windows of zero transmission presented in 
Sec.~\ref{pn}.

We note that $T_{\rm min}$ obtained in this simple model is due 
to the local band energy mismatch close to the contact regions. It should 
not be confused with the minimum transmission in disordered graphene 
systems addressed in Sec.~\ref{subsec:puddle_disorder}.

\subsection{Focusing effects in a two Gaussian puddle geometry}\label{twogaussians}

\begin{figure}[!htbp]
\centering
\includegraphics[width=\columnwidth]{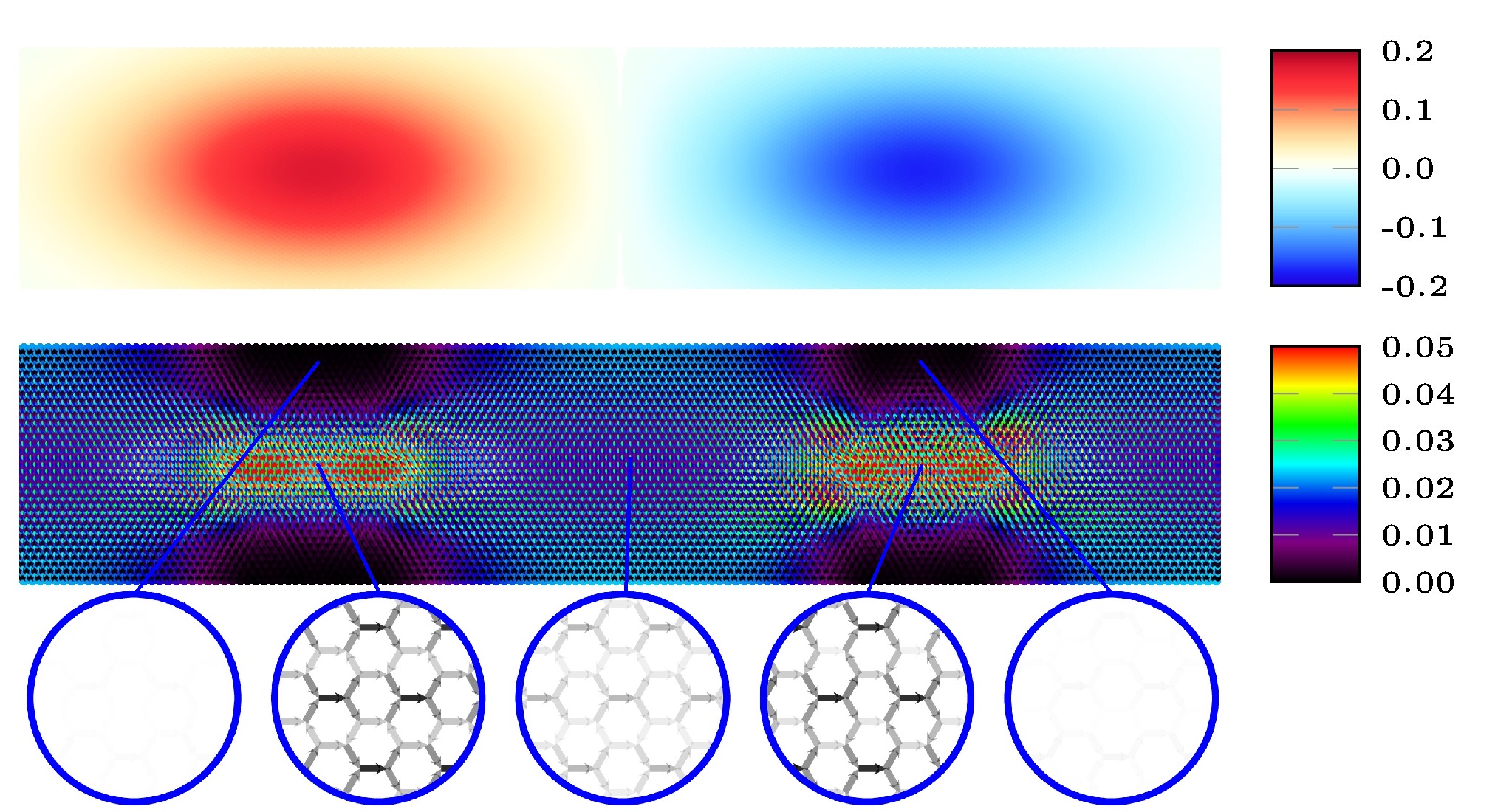}
\caption{(Color online) Top: Potential profile $V({\bf r})$ for $M=52$. 
The energy scale in the right is in units of $t$.
Middle: Corresponding local transmission at $E_F\approx 0.15E_1$. 
Bottom: Zoom of the local transmission at selected areas. 
}
\label{fig:030}
\end{figure}

Let us now study the effect of a potential variation along the transverse direction. 
Specifically, we analyze the transmission in a graphene ribbon with two Gaussians 
charge puddles. We set $d_y = 0.6Ma_0$, $d_x =0.24L$ and place the 
Gaussians potentials at $(L/4,W/2)$ and $(3L/4,W/2)$.
The remaining parameters are the same as in Sec.~\ref{gaussianpn} with $U=t$.
The potential profile $V({\bf r})$ is illustrated by Fig.~\ref{fig:030} (top panel).

The middle panel of Fig.~\ref{fig:030} shows the local transmission profile at 
$E_F=0.15E_1$.
Near the source (left) and the drain (right) the local transmission is almost evenly 
distributed across the system width. 
At the \textit{pn} interface region, $x\approx L/2$, the behaviour is similar.
In contrast, near the center of both positive and negatively charged puddles, 
corresponding to the most doped regions of the system, the transmission is 
largely enhanced. This implies in a strong suppression of the current close to the 
edges, due to current conservation along different ribbon cross sections.

In summary, the Gaussian potentials not only favor the electronic propagation, but 
also focus the transmission on the highly doped areas. 
This effect can be interpreted in terms of the picture discussed in the previous section.
Close to the center of the puddles, the potential $V({\bf r})$ shifts the local dispersion 
relation. ``Local" transmission modes are opened if $|V({\bf r}_c)| > E^*_n$. Note that 
here the threshold energies $E^*_n$ are related to the puddle width $a$, rather then to the 
system width $W$.

\begin{figure}[!htbp]
\centering
\includegraphics[width=0.95\columnwidth]{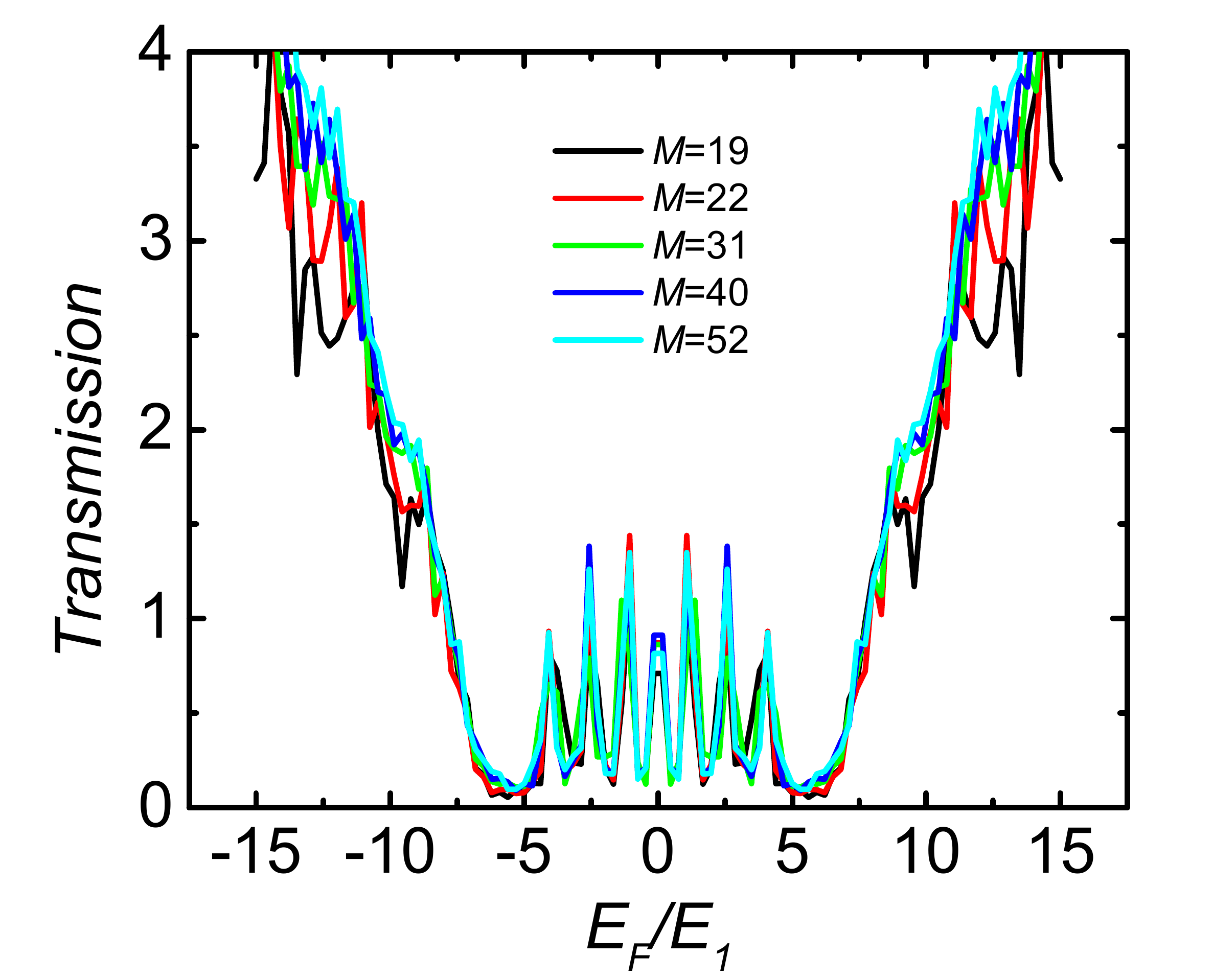}
\caption{(Color online) Transmission as a function of the electronic energy $E_F$ in units 
of $E_1$ for armchair edge graphene ribbons of different widths $M$. 
}
\label{fig:disorder_collapse_030}
\end{figure}

As in the previous case, we find that the transmissions $T(E_F)$  for different 
widths $W$ collapse very nicely to a single curve by scaling $\epsilon=E_F/E_1$, 
particularly for the energy window where $|\epsilon|<5$, see Fig.~\ref{fig:disorder_collapse_030}.
We expect a similar result if we scale $E_F$ by $E_1^*$, since in our model the ratio 
between the puddle size $a$ and the system width $W$ does not change with $M$.
These observations suggest that by proper scaling one can address systems with realistic sizes.

At a first inspection, $T(E_F)$ shown in Figs.~\ref{fig:disorder_collapse_999_inset} 
and \ref{fig:disorder_collapse_030} look similar.
A more detailed analysis indicates that in the present case: 
(i) The value of the transmission minimum is smaller.
(ii) The interference pattern covers a smaller energy window. 
We speculate that those results are due to the smaller doping of the areas near the edges.
The ribbon accommodates a smaller number of propagating states, compared to the 
\textit{pn} junction case, so that the total transmission through a ribbon cross section is 
smaller. Since the total doping of the ribbon is smaller, the effective scattering potential 
that determines the energy window of the interference pattern is also smaller.  

\subsection{Disordered charge puddles}
\label{subsec:puddle_disorder}

In this subsection we study the case of randomly distributed charge 
puddles, that are ubiquitous in graphene samples 
\cite{castroneto09,Katsnelson12book,Martin08,Zhang2009,Xue11}.

\begin{figure}[!htbp]
\centering
\includegraphics[width=\columnwidth]{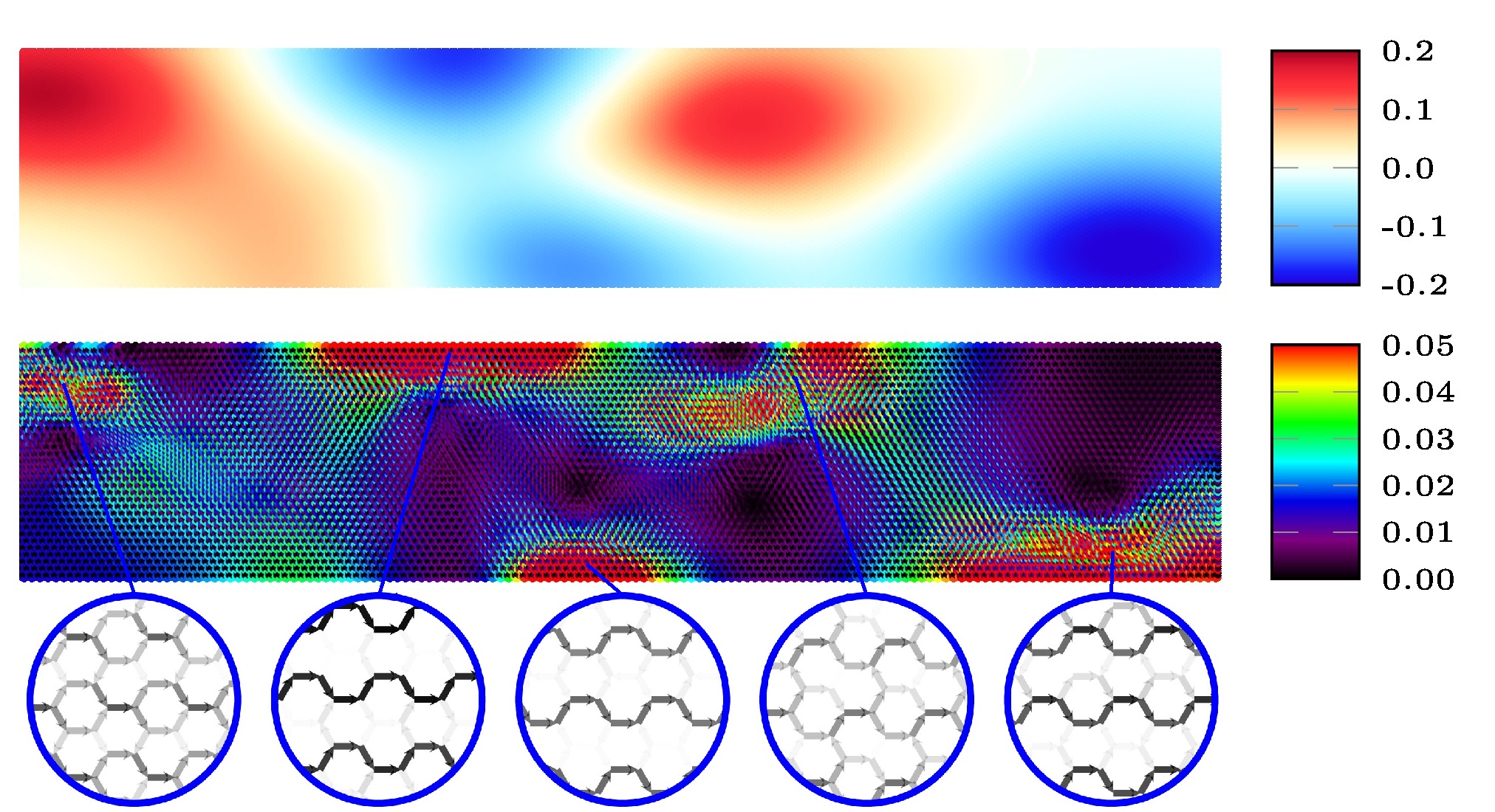}
\caption{(Color online) Top: Random potential profile $V({\bf r})$ realization 
(see text for details).  
The energy scale in the right is in units of $t$.
Middle: Corresponding dimensionless local current density at energy 
$E_F\approx 2.31E_1 = 0.04t$.
Bottom: Zoom of selected areas showing the local transmission in details.
}
\label{fig:randfig2}
\end{figure}

\begin{figure}[!htbp]
\centering
\includegraphics[width=\columnwidth]{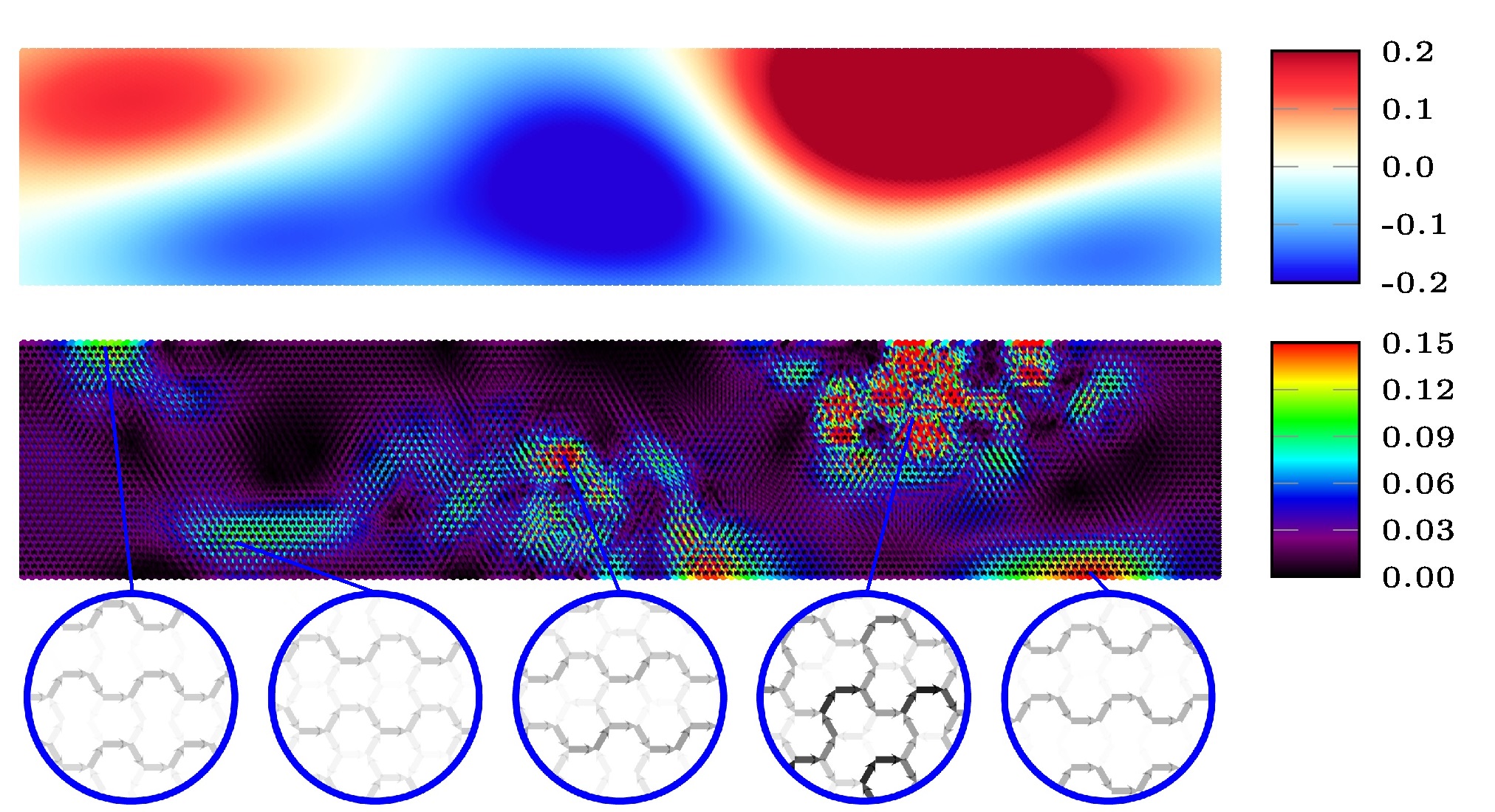}
\caption{(Color online) Same as in Fig.~\ref{fig:randfig2} for a different random potential 
realization.
}
\label{fig:randfig8}
\end{figure}

We analyze two limiting cases, namely, small and large charge puddles, as 
compared with $W$. The corresponding $V({\bf r})$ are shown in the top 
panels of Figs.~\ref{fig:randfig2} (small puddles) and \ref{fig:randfig8} (large puddles). 
The random potential is generated according to Eq.~\eqref{potential} with $N_G=8$,
$d_x=62a_0$ and $d_y=31a_0$ for a system with dimensions $M=52$ ($W=12.5$nm) 
and $N=604$ ($L=62.8$nm).
In both case we set $V=0.2t=11.44E_1$.

The middle and bottom panels of Figs.~\ref{fig:randfig2} and \ref{fig:randfig8} 
show that, as in the previous subsection, the local transmission is focused on the
maximally \textit{n} and \textit{p} doped areas.
The ``large" puddles illustrated in Fig.~\ref{fig:randfig8} induce higher local 
currents than the ``small" ones corresponding to Fig.~\ref{fig:randfig2} (see scales). 

The results indicate that the local transmission depends strongly on the potential 
landscape, since the number of propagating modes increases with both puddle 
size $a$ and local doping $\delta n({\bf r}) \equiv n({\bf r},V\neq 0) - n({\bf r},V=0)$. 
In the limit of $a (\delta n)^{1/2} \gg 1$, the random resistor model put forward in 
Ref.~\onlinecite{Cheianov07} estimates the conductivity at the CNP to be 
$\sigma_{\rm min} \approx (e^2/\hbar)(a^2 \delta n)^{0.41}$, where
$a$ and $\delta n$ are defined by the correlation function 
{
$\langle \delta n(\mathbf r) \delta n(\mathbf r') \rangle \equiv \delta n^2 
F\left(|{\bf r}-{\bf r}'|/a\right)$
}
\cite{Cheianov07}.
The model is semiclassical and does not include interference effects. 
Despite this limitations, $\sigma_{\rm min}$ is in qualitative agreement
with our numerical findings we discuss next, namely, that the transmission 
near the CNP is larger for the ``large" charge puddle case than for the 
``small" puddles one.

\begin{figure}[!htbp]
	\centering
		\includegraphics[width=0.95\columnwidth]{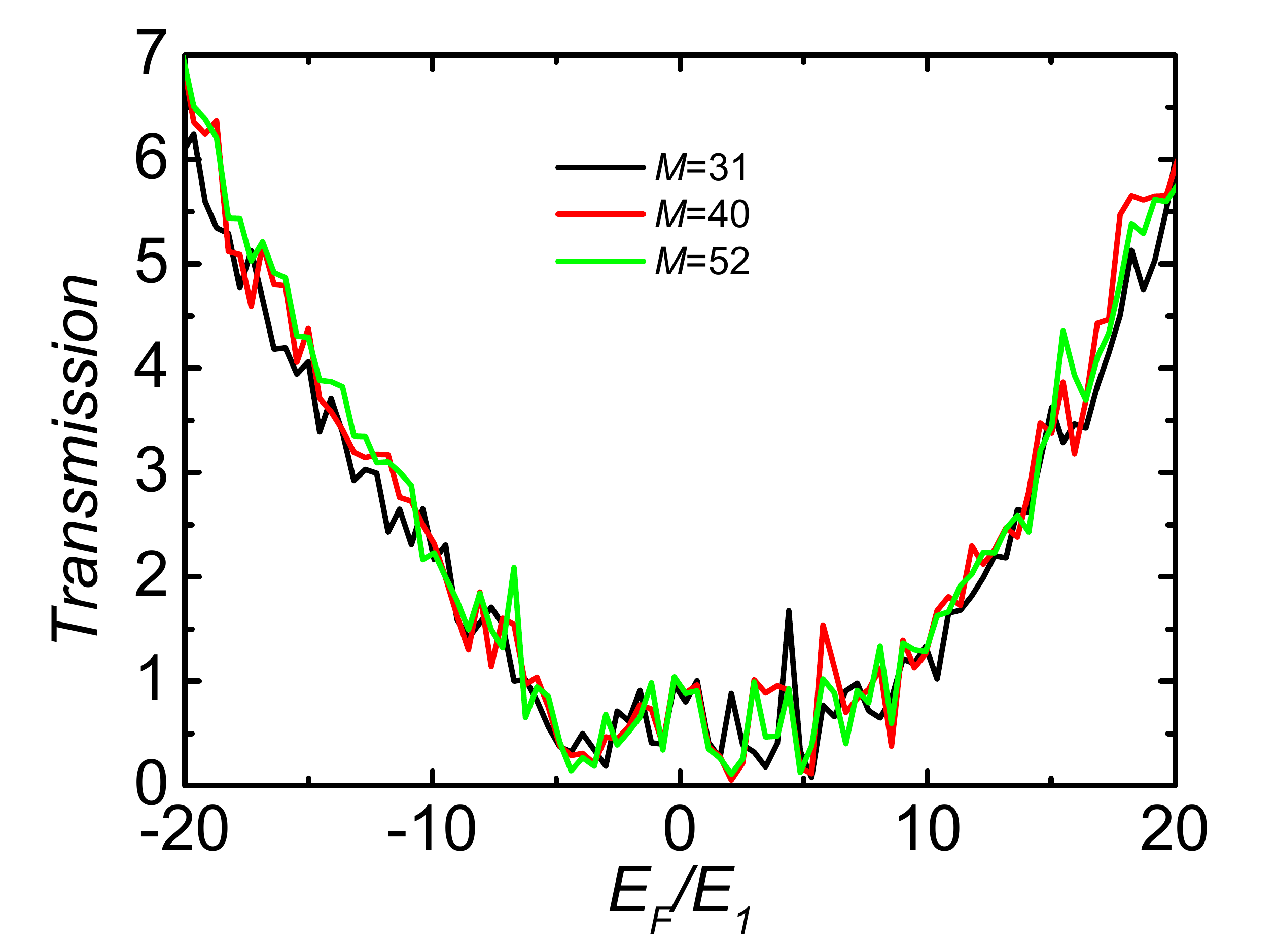}
	\caption{(Color online) Transmission as a function of the electronic energy in units of $E_1$. We show the results for $M=52$ with the potential profile in Fig.~\ref{fig:randfig2} and for $M=31,40$ using a similar potential profile scaled down to smaller sizes keeping the aspect ratio.
	}
	\label{fig:random_puddles_collapse_fig2}
\end{figure}
%
\begin{figure}[!htbp]
	\centering
		\includegraphics[width=0.95\columnwidth]{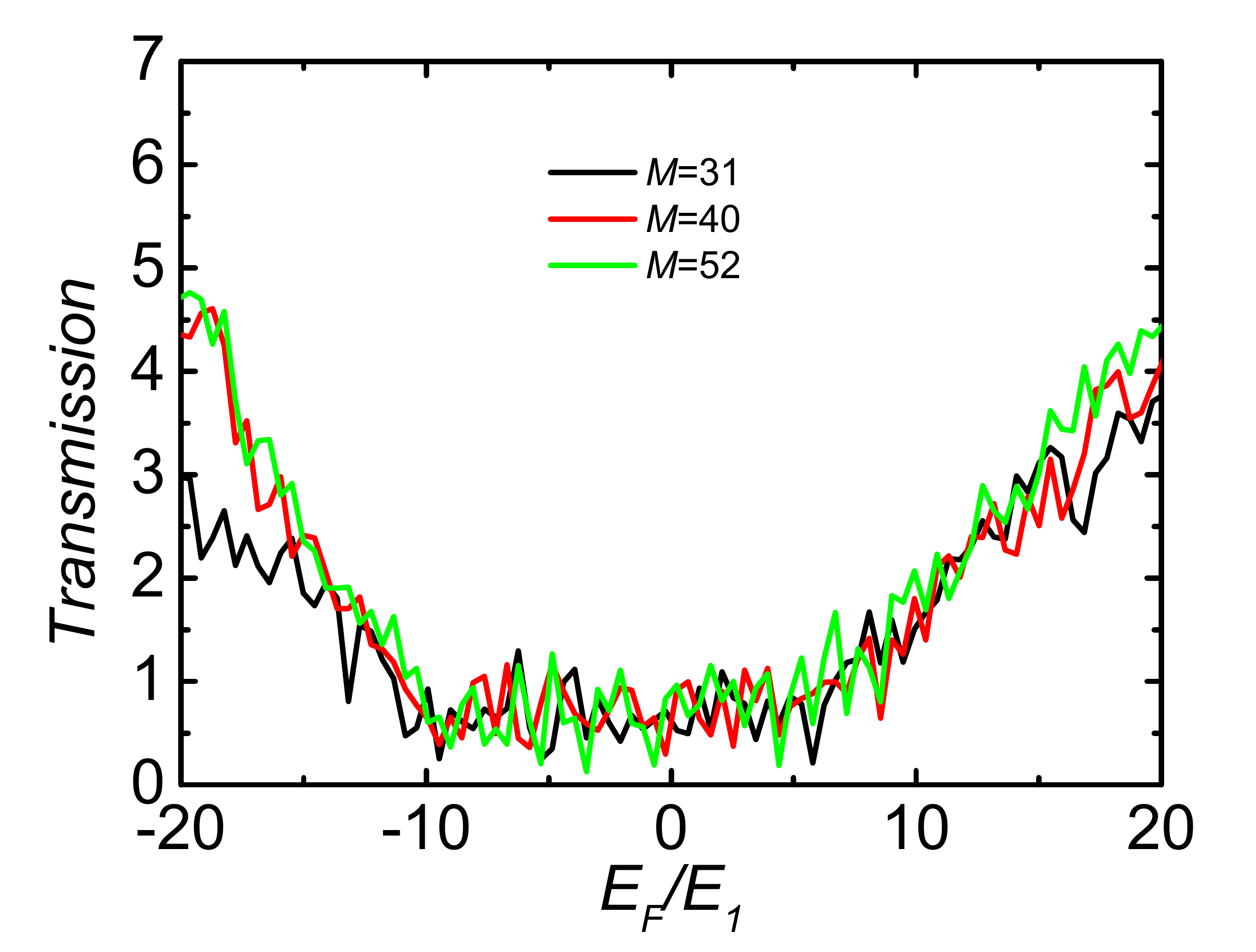}
	\caption{(Color online) Same as in Fig.~\ref{fig:random_puddles_collapse_fig2} using the potential profile in Fig.~\ref{fig:randfig8}.	}
	\label{fig:random_puddles_collapse_fig8}
\end{figure}

Figures \ref{fig:random_puddles_collapse_fig2} and \ref{fig:random_puddles_collapse_fig8} show
the total transmission corresponding to the potential profiles presented by Figs.~\ref{fig:randfig2} 
and \ref{fig:randfig8}, respectively. Here, by expressing the energy as $\epsilon = 
E_F/E_1$, the transmissions for different system sizes no longer collapse into a single curve.  
Nonetheless, in all studied cases $T(\epsilon)$ shows a similar average behavior and 
fluctuations reminiscent of the universal conductance fluctuations (UCF) ubiquitous in 
disordered mesoscopic systems.

Around the CNP ($|E/E_1|\alt 10$) the number of open modes depend strongly on the 
typical puddle size: The larger the puddles the smaller $E^*_n$. Hence, one expects to 
observe an enhanced transmission for the ``large" puddle case. 
Away from the CNP ($|E_F/E_1| \agt 10$), Figs.~\ref{fig:random_puddles_collapse_fig8} 
and \ref{fig:random_puddles_collapse_fig2} show that the transmission for the ``large" 
puddles case is smaller than that for the ``small" ones.
We interpret this feature as follows: Here the number of open modes is dictated by $E_F$. 
Stronger disorder potential fluctuations are more effective to mix different modes and to 
favor backscattering, giving rise to a smaller transmission.
Thus, disordered puddles are detrimental to the transport as one approaches 
the semiclassical regime, but they enhance the transmission minimum around 
the CNP.
We note that here the transmission minimum survives even in the absence of 
evanescent modes \cite{Tworzydlo06}.

We also compare the transmission for $U=0$ and $U=t$. The results are qualitatively similar. 
We conclude  the interaction $U$ does not play a central in the results presented in this paper.
Our calculations (not shown here) demonstrate that the interaction corrects the overestimation 
of the onsite electronic density, but has little impact on the local or total transmissions through 
the systems.

We make connection with experiments by estimating the typical values of $E_1^*$ 
for realistic size samples.
We find that graphene charge puddles with sizes $a \approx 20 \cdots 50$~nm
correspond to $E_1^* \approx 10 \dots 30$ meV.
Typical graphene on silicon oxide samples \cite{Xue11} show $\delta V/E_1^*>1$. Hence,
charge puddle disorder enhances the conductivity and guarantees a non-vanishing 
conductivity minimum at the CNP, independent of the contribution due to evanescent modes.
For graphene on hBN, where $\delta V \approx 5$ meV \cite{Xue11}, only a small fraction of
puddles meet the criterion $V_{\rm loc}(\mathbf r)/E_1^* \agt 1$. In this situation, charge puddle 
fluctuations assisted transport is very unfavorable. Hence, for graphene flakes on hBN 
with aspect ratios $L/W > 3$, where evanescent modes contribute very little to the 
transmission, one expects the conductivity at the CNP to be strongly suppressed. 
We believe that this scenario is consistent with the experimental report \cite{Amet13} of 
an insulator behavior of single-layer graphene on hBN  at the neutrality point.

\section{Conclusions}
\label{sec:conclusions}

We studied the effect of charge puddles in the transmission minimum of single layer graphene stripes by 
means of a microscopic model based on a spin resolved tight-binding Hamiltonian 
including electron-electron interactions via a Hubbard mean field term. 
To understand the conductivity at the CNP and scale up our results to experimental size 
samples, we used the recursive Green's functions technique to obtain the transmission through 
semiconductor graphene strips with armchair edge. The charge puddles are modeled by a 
local Gaussian disordered potential.

First we studied pristine graphene systems with a smooth \textit{pn} junction. This simple model 
shows that the onset of the transmission minimum at the CNP occurs for potential strengths $V$ 
larger than the threshold energy $E_1$ to open the first conducting transversal mode of the system.
The transmission near the CNP is robust against smooth changes in the potential along the 
propagation direction and it does not depend much on whether the doping is \textit{n}- or \textit{p}-type. 
We showed that all transmission features around the CNP can be explained in 
by Klein tunneling and by Fabry-Perot interference due to the mismatch of the 
wave functions at the graphene-contacts
interface.

Next, we studied the transmission through a system with two charge puddles separated by a 
smooth {\it pn} interface.
In this setting, we find that the overall total transmission decreases and the local 
transmission is focused around the maximally \textit{p} or \textit{n} doped regions, corresponding to the 
centers of the puddles.
We demonstrated that by using $E_1$ (or $E^*_1$, see text) as the energy unit, the transport 
properties around the CNP become independent of the system size.
This powerful result allows us to address realistic sized systems by scaling up our model
calculations obtained for small systems, whose sizes are imposed by computational limitations.

Finally, we also modeled disordered charge puddles distributions showing that, 
depending on the puddles sizes, there is a non-vanishing average transmission 
minimum around the CNP with fluctuations similar to UCF. The numerical results 
can be qualitatively explained by Klein tunneling 
at the {\it pn}-interfaces formed at the puddles interface and the enhanced (focussed) local transmission at the maximally doped areas. 

Our results show that, for graphene on silicon oxide, the local chemical fluctuations 
\cite{Xue11,Deshpande11} are sufficiently large  to explain a non-vanishing conductivity 
minimum at the charge neutrality point $\sigma_{\rm CNP}$ in terms of charge puddle 
disorder assisted transport.
On the other hand, in typical graphene samples on hBN \cite{Xue11}, only a small 
fraction of puddles show $V_{\rm loc}(\mathbf r)/E_1^* \agt 1$. 
In this case, unless compensated by contributions from evanescent modes, one expects
a strong suppression of $\sigma_{\rm CNP}$.
This scenario is consistent with the recent experimental report \cite{Amet13} 
of an insulator behavior of  $\sigma_{\rm CNP}$ in graphene on hBN samples with
an aspect ratio $L/W \agt 3$.

In summary, this study separates the contribution of evanescent modes from that of charge puddles in the transport properties of graphene strips close to the CNP. We found that the presence of electron and hole puddles in graphene enhances the electronic transmission at the CNP depending their size and charge,
represented in our model by $a$ and $V$. We argue that our findings provide a scenario to explain transport experiments in graphene deposited on both SiO$_2$ \cite{Geim07,Tan07} and hBN substrates \cite{Amet13}.  

\begin{acknowledgments}
This work is supported by the Brazilian funding agencies CAPES, CNPq, and FAPERJ.
\end{acknowledgments}

\bibliography{graphene_charge_puddles}

\end{document}